\documentstyle[12pt]{article}
\begin{document}
\title{SPACE TIME AS A RANDOM HEAP}
\author{B.G. Sidharth$^*$\\
B.M. Birla Science Centre, Hyderabad 500 063 (India)}
\date{}
\maketitle
\footnotetext{$^*$E-mail:birlasc@hd1.vsnl.net.in}
\begin{abstract}
In this paper, we demonstrate how space-time is, rather than a differentiable
manifold, a Random Heap, and how this ties up with fractal dimension 2 of
a Quantum Mechanical path. In this light, we can see that there is a
harmonious convergence between the stochastic approach of Nelson and the
de Broglie-Bohm approach. These considerations are shown to lead to the
emergence of special relativity and Quantum Mechanics.
\end{abstract}
\section{Introduction}
Space time has generally been taken to be a differential manifold
with an Euclidean (Galilean) or Minkowskian or Riemannian character.
Though the Heisenberg Principle in Quantum Theory forbids arbitrarily small
space time intervals, the above continuum character with space time
points has been taken for granted. Indeed it has been suggested by Snyder,
Lee and others\cite{r1,r2,r3,r4} that the infinities which plague
Quantum Field Theory are symptomatic of the fact that space time has a
granular or discrete character. This has lead to a consideration of extended
particles, as against point particles of conventional theory. Wheeler's
space time foam and strings\cite{r5,r6} are in this class, with a
minimum cut off at the Planck scale. This has also lead to a review of the
conventional concept of a rigid background space time. More recently
\cite{r4,r7,r8}, it has been pointed out that it is possible to give
a stochastic underpinning to space time and physical laws. This is in the
spirit of what Wheeler calls, "Law without Law"\cite{r9}.\\
In the light of the above remarks, we now point out that space time is
fractal and can be considered to be what may be termed a Brownian or Random
Heap. This would be at the root of Quantum behaviour.
\section{The Random Heap}
What distinguishes Quantum Theory from Classical Physics is the role of the
resolution of the observer or observing apparatus. Indeed as noted by
Abbot and Wise\cite{r10}, in this respect the situation is similar to everywhere
continuous but non differenciable curves, the fractals of Mandelbrot
\cite{r11}. This again is tied up with the Random Walk or Brownian character
of the Quantum path as noted by Sornette and others\cite{r12}: At scales larger than
the Compton wavelength but smaller than the de Broglie wavelength, the
Quantum paths have the fractal dimension 2 of Brownian paths (cf. also
Nottale,\cite{r13}). It must be noticed that both these length scales involve
the mass. It will be seen below that this is quite meaningful.\\
Two important characteristics of the Compton wavelength have to be re-emphasized
(Cf.\cite{r4,r8}): On the one hand with a minimum space time cut off at the
Compton wavelength, we can recover by a simple coordinate shift the Dirac
structure for the equation of the electron, including the spin half. In this
sense the spin half, which is purely Quantum Mechanical is symptomatic
of the minimum space time cut off, as is also suggested by the Zitterbewegung
interpretation of Dirac (in terms of the Uncertainity Principle), Hestenes
and others\cite{r14,r15,r16}. The Zitterbewegung is symptomatic
of the fact that by the Heisenberg Uncertainity Principle, Physics begins
only after an averaging over the minimum space time intervals. This is also
suggested by stochastic models of Quantum Mechanics, both non relativistic
and relativistic\cite{r17,r18,r19,r4}.\\
The other aspect is that the Compton wavelength $l$ of a typical elementary
particle, the pion is given by the well known empirical "Large Number" relation
\begin{equation}
R \approx \sqrt{N} l\label{e1}
\end{equation}
where $R \sim 10^{28}cm$ is the radius of the universe and $N \sim 10^{80}cm$
is the number of elementary particles in the universe.
As pointed out,\cite{r4,r8} (\ref{e1}) is also the Random Walk relation.
Infact (\ref{e1}) and a similar equation for the Compton time in terms of
the age of the universe, viz.,
\begin{equation}
T \approx \sqrt{N} \tau\label{e2}
\end{equation}
were the starting point for a unified scheme for physical interactions and
indeed a cosmology consistent with observation. It was pointed out
\cite{r20} that in the spirit of Wheeler's travelling
salesman's "practical man's minimum" length that the Compton scale plays
such a role, and that space time is like Richardson's delineation of a
jagged coastline with a thick brush, the thickness of the brush being
comparable to the Compton scale.\\
Space time, rather than being a smooth continuum, is more like a fractal
Brownian curve. To analyse this further, we observe that space time given
by $R$ and $T$ of (\ref{e1}) and (\ref{e2}) represents a measure of dispersion
in a normal distribution. Indeed if we have a large collection of $N$ events
(or steps) of length $l$ or $\tau$, forming a normal distribution, then the
dispersion $\sigma$ is given by precisely the relation (\ref{e1}) or
(\ref{e2}).\\
The significance of this is brought out by the fact that the universe
is a collection of $N$ elementary particles, infact typically pions of size
$l$, as seen above (Cf. ref.\cite{r21}). We consider space time not as an apriori
container but as a Gaussian collection of these particles. It is a Random
Heap. At this stage, we do not even need the concept of a continuum.\\
In this scheme the probability distribution has a width or dispersion
$\sim \frac{1}{\sqrt{N}}$ (Cf. ref.\cite{r22}), that is the fluctuation
(or dispersion) in the number of particles $\sim \sqrt{N}$. This immediately
leads to equations (\ref{e1}) and (\ref{e2}). Moreover it leads to a completely
consistent cosmology as pointed out earlier\cite{r23} which explains how all the
so called Large Number coincidences and also Weinberg's "mysterious" formula
relating the pion mass to the Hubble constant,
\begin{equation}
m^3 = \left(\frac{H\hbar^2}{Gc}\right)\label{e3}
\end{equation}
far from being empirical, follow as a consequence of the theory,
while predicting an ever expanding universe as is confirmed by latest
independent observations.\\
It must be emphasized that equations (\ref{e1}) and (\ref{e3}) in particular
bring out a holistic or Machian feature in which the large scale universe
and the micro world are inextricably tied up, as against the usual differential
view. This is infact inescapable if we are to consider a Brownian Heap.\\

\section{Stochastic Considerations}
It was observed in Section 2 that the cut off length for fractal
behaviour depends on the mass, via the de Broglie or Compton wavelength. The
de Broglie wavelength is the non-relativistic version of the Compton wavelength.\\
Indeed it has been shown in detail\cite{r16,r24} that it is the Zitterbewegung
or self-interaction effects within the minimum cut off Compton wavelength that
indeed give rise to the inertial mass. So the appearance of mass in the minimum
cut off Compton (or de Broglie) scale is quite natural.\\
With this background we observe that it has been pointed out by Nottale\cite{r13}
and others, though from a
different standpoint, that the fractal nature and a stochastic underpinning
are interrelated: for times less than the Compton (or de Broglie) wavelength,
time runs backwards, corresponding to Nelson's double Wiener process\cite{r25}.
This leads to the complex wave function of Quantum Mechanics.\\
This is also the Zitterbewegung within the Compton scales as discussed
elsewhere (Cf.ref.\cite{r16}).\\
Infact we have here a harmonious and meaningful convergence of Nelson's stochastic approach
and the de Broglie-Bohm approach, if we recognize the minimum space time
cut offs, which again as pointed out, have a Brownian underpinning. This can
be seen briefly as follows.\\
Let us consider the motion of a particle with position given by $x(t)$, subject
to random correction given by, as in the usual theory, (Cf.\cite{r25,r8}),
$$|\Delta x| = \sqrt{<\Delta x^2 >} \approx \nu \sqrt{\Delta t},$$
the diffusion constant $\nu$ being given by
$$\nu = \hbar/m,$$
and being related to the mean free path by
\begin{equation}
\nu \approx l v\label{e4}
\end{equation}
We can then proceed to the Fokker-Planck equation.\\
In Nelson's derivation, the Schrodinger wave function, exactly as in the
de Broglie-Bohm approach, is decomposed as (Cf. also\cite{r26})
$$\psi = \rho e^{\imath s/\hbar},$$
which in both cases leads to the well known Hamilton-Jacobi type equation
\begin{equation}
\frac{\partial S}{\partial t} = -\frac{1}{2m} (\partial S)^2 + V +Q,\label{e5}
\end{equation}
where
$$Q = \frac{\hbar^2}{2m} \frac{\nabla^2 \sqrt{\rho}}{\sqrt{\rho}}$$
As pointed out by Nottale\cite{r26}, the complex feature above disappears if the
fractal or non-differential character is not present, (that is, the forward and
backward derivates are equal): Indeed the fractal
dimension 2 leads to the real coordinate becoming complex. What distinguishes
Quantum Mechanics is the adhoc feature, the diffusion constant $\nu$ of
(\ref{e4}) in Nelson's theory and the "Quantum potential" $Q$ of (\ref{e5}).\\
However, once we recognize the minimum cut off space-time intervals, these
adhoc features become quite meaningful: Equation (\ref{e4}) gives us the
Compton wavelength and the $Q$ of the equation (\ref{e5}), as described in
reference\cite{r16} gives the inertial mass and energy of "quantized vortices"
with the same Compton scale extent.\\
The following will throw further insight on the foregoing considerations.
Let us start with the Langevin equation in the absense of external forces,
$$m \frac{dv}{dt} = -\alpha v + F'(t)$$
where the coefficient of the frictional force is given by Stokes's Law
(cf.\cite{r27})
$$\alpha = 6\pi \eta a$$
This then leads to two cases.\\
Case (i):\\
For $t$, there is a cut off time $\tau$. This is so because
$$\frac{n}{\alpha} = \frac{m}{\eta a},$$
so that, as per Stokes's Law, as
$$\eta = \frac{mc}{a^2}$$
we get
$$\tau \sim \frac{ma^2}{mca} = \frac{a}{c},$$
that is $\tau$ is the Compton time.\\
The expression for $\eta$ which follows from the fact that
$$F_x = \eta (\Delta s) \frac{dv}{dz} = m\dot v = \eta \frac{a^2}{c} \dot v ,$$
shows that the intertial mass is a type of viscosity of the background
ZPF reminiscent of the work of Rueda and Haisch\cite{r28} and similar to the
Compton wavelength mass referred to earlier (cf.also ref.\cite{r21}).
To sum up case (i), for a cut off $\tau$, the stochastic form leads us back to the
minimum space time intervals $\sim$ Compton scale.\\
To push these small scale considerations further, we have, using the
Beckenstein radiation equation\cite{r29},
$$t \equiv \tau = \frac{G^2m^3}{\hbar c^4} = \frac{m}{\eta a} = \frac{a}{c}$$
which gives
$$a = \frac{\hbar}{mc} \quad \mbox{if} \quad \frac{Gm}{c^2} = a$$
In other words the Compton wavelength equals the Schwarzchild radius, which
automatically gives us the Planck mass.\\
On the other hand if we work with $t \geq \tau$ we get
$$ac = \frac{2kT}{\eta a}$$
whence
$$kT \sim mc^2,$$
which is the Hagedorn formula for Hadrons\cite{r30}.\\
Thus both the Planck scale and the Compton wavelength Hadron scale
considerations follow meaningfully.\\
Case (ii):\\
If there is no cut off time $\tau$, as is known, we get back,
$$\Delta x = \nu \sqrt{\Delta t}$$
and thence Nelson's derivation of the non relativistic Schrodinger
equation.\\
\section{Space Time}
As remarked in the previous section, the fact that forward and backward time
derivatives in the double Wiener process do not cancel leads to a complex
velocity (cf.\cite{r26}), $V-\imath U$. That is, the usual space coordinate
$x$ (in one dimension for simplicity) is replaced by a coordinate like
$x+\imath x'$, where $x'$ is a non constant function of time. We will now
show that it is possible to consistently take $x' = ct$.\\
Let us take the simplest choice for $x'$, viz., $x' = \lambda t$. Then the
imaginary part of the complex velocity is given by $U = \lambda t$. Then
we have (cf.\cite{r26}),
$$U = \nu \frac{d}{dx}(ln\rho) = \lambda$$
where $\nu$ and $\rho$ have been defined in the equations leading to (\ref{e4})
and (\ref{e5}). We thus have, $\rho = e^{\gamma x}$, where $\gamma = \lambda/\nu$
and the quantum potential of (\ref{e5}) is given by
\begin{equation}
Q = \frac{\hbar^2}{2m}\cdot \quad \gamma^2\label{e6}
\end{equation}
We have already remarked in the previous section that in this stochastic
formulation with Compton wavelength cut off, $Q$ turns out to be the inertial energy
$mc^2$ (cf.\cite{r16}). It then follows from (\ref{e6}) that $\lambda = c$.\\
In other words it is in the above stochastic (and fractal) formulation that we
see the emergence of the space time coordinates $(x,\imath ct$). If we now
generalise to three spatial dimensions, then as is well known\cite{r31}, we get
the quarternion formulation with the Pauli spin matrices giving the purely
Quantum Mechanical spin half of Dirac. On the other hand, the above fractal
formulation with minimum space time cut offs has been shown to lead
independently to the Dirac equation\cite{r4} as remarked earlier. Thus the origin of special
relativity, inertial mass and the Quantum Mechanical spin half is the minimum
space time cut offs. (Indeed it can be shown that a minimum space time cut off
leads to Special Relativity\cite{r8,r19}.
\section{Comments}
(i) The fact that the Quantum Mechanical wave function is complex, which indeed
is one of the distinguishing features, is directly related to the above minimum
space time cut offs and fractal considerations. We must also rememeber that
the Quantum Mechanical wave function contains as much information of the
system under consideration as possible. The values of the time derivative
of the wave function cannot therefore appear as initial conditions.\\
Interestingly it can
be shown that this consideration together with the requirement of causality
demands that the wave function be complex\cite{r32}. If the wave function
were to be real, then it is well known that we have a stationary situation
in the Quantum picture. In Classical Mechanics, in contrast the dynamics
is in the second time derivative, and both a quantity like the position
and its derivative, that is velocity or momentum are required for initial
conditions. That is why in Classical theory we do not require complex
quantities.\\
(ii) It may be mentioned, that the holistic feature is contained
in (\ref{e1}) and (\ref{e3}). In this spirit, starting from a universal set - for example
the universe of galaxies - and considering a cascade of subsets, it is
possible to see the origin of a metric (Cf.\cite{r33} for details), rather
than start with particles and consider larger and larger sets.\\
(iii) The considerations of Section 3 and Section 4 show how Nelson's stochastic
formulation and the de Broglie-Bohm approach converge. Both these approaches
however have been non relativistic and have had some unsatisfactory adhoc
features. For example the diffusion constant which appears in (\ref{e4})
or the non local Quantum potential which appears in (\ref{e5}). These features
however become meaningful once we take into account the Compton scale cut
off and the Zitterbewegung effects, as described in the previous section.
It is also this stochastic or fractal feature which ensures that the underlying
theory has no hidden variables (Cf.refs.\cite{r13} and \cite{r20}).

\end{document}